\documentclass{article}

\title{Conformal Mapping of Relativistic Quantum Bound Systems to Eliminate Potential Fields}
\author{Robert Ducharme}

\begin{document}
\maketitle

\centerline{151 Fairhills Dr., Ypsilanti, MI 48197}
\centerline{E-mail: ducharme01@comcast.net}

\begin{abstract}
In two recent papers, an isometric conformal transformation has been introduced that eliminates potential interaction terms from the Schr\"{o}dinger equation for central potential problems. The method has been demonstrated for both the hydrogen atom and three-dimensional harmonic oscillator. Here, it is shown that the same transformation technique can also be applied to central potential problems formulated using the Klein-Gordon equation.  
\end{abstract}

\section{Introduction}
Conformal mapping \cite{ZN} is a coordinate transformation technique that has found numerous practical applications in science and engineering. Most of these applications are for two-dimensional problems but the scope of the technique is not limited to two-dimensions. Liouville's theorem \cite{DEB} in fact shows that higher dimensional conformal maps are possible but must be composed of translations, similarities, orthogonal transformations and inversions. In two recent papers, an isometric conformal transformation has been used to simplify the Schr\"{o}dinger equation for the three-dimensional harmonic oscillator \cite{RJD1} and the hydrogen atom \cite{RJD2}. The purpose of this paper is to show the same transformation technique can also be similarly applied to central potential problems formulated using the Klein-Gordon equation.

It is convenient to consider a single particle of energy E and a spatial displacement $x_i$ $(i=1,2,3)$ from a potential source. In the rest frame of this system, both the particle and the source may be assumed to exist at the same time t. The goal of section 2 of this paper is to introduce an isometric conformal mapping of the form $z_i= x_i,s=t-\imath \hbar \mathcal{F}(|x_i|,E)$ where $\mathcal{F}$ is a real function, $\hbar$ is Planck's constant divided by 2$\pi$ and $\imath=\sqrt{-1}$. It is clear from inspection that this passive transformation does no more than introduce an imaginary shift along the time-axis of the two related coordinate systems.

It is convenient to write the complex conjugate form of the $(z_i,s)$-coordinates as $(z_i^*,s^*)$ even though $z_i^*=z_i$ since $z_i^*$ and $z_i$ still belong to different coordinate systems. This distinction is shown to be most evident in the computation of the partial derivatives $\partial / \partial z_i$ and $\partial / \partial z_i^*$ from the chain rule of partial differentiation as these evaluate differently in their conjugate coordinate systems. One further topic to be introduced in section 2 is the Cauchy-Riemann equations that are necessary to determine if a function transformed into the $(z_i,s)$-coordinate system has well defined partial derivatives. 

The Klein-Gordon equation for the harmonic oscillator is presented in section 3 alongside a complete set of eigensolutions. It is shown that both these results simplify in terms of $(z_i,s)$-coordinates and that the eigensolutions are holomorphic. It is of interest that the harmonic oscillator potential is eliminated from the Klein-Gordon equation in  $(z_i,s)$-coordinates. It is also convenient that the lowering and raising operators for the harmonic oscillator are proportional to the partial derivatives $\partial / \partial z_i$ and $\partial / \partial z_i^*$ respectively.

In section 4, the Klein-Gordon equation for a charged particle in a Coulomb field is mapped into $(z_i,s)$-coordinates. As in the harmonic oscillator case, any reference to the potential field is eliminated from the mathematical description of the problem through the mapping. 

\section{Conformal Mapping}
The task ahead is to present an isometric conformal transformation relating a real $(x_i,t)$-coordinate system and a complex $(z_i,s)$ coordinate system. This mapping is intended for application to a particle of mass $m$ and total energy $E$ at a radial separation $r$ from a point source. It is convenient to express it in the form
\begin{equation} \label{eq: conftrans1}
z_{i} = x_{i}, \quad s = t - \imath \frac{\hbar}{E} \left[ a\ln(r) + \left(\frac{r}{b} \right)^\lambda \right]
\end{equation}
where $\lambda=1$ for the hydrogen atom, $\lambda=2$ for the harmonic oscillator. The form of the dimensionless quantity $a$ and scale length $b$ must be determined for each specific problem. It is notable the logarithmic term is absent in previous papers that treat non-relativistic problems. It is introduced here for the relativistic treatment of a charged particle in a Coulomb field.

In the $(x_i,t)$-coordinate system, the particle has a spatial displacement $x_i$ from the potential source but shares the same world time t as the source. Similarly, in the $(z_i,s)$-coordinate system, the particle has a displacement $z_i = x_i$ from the source but shares the same complex time s. The isometric nature of the transformation therefore follows from the result $|z_i|=|x_i|$. It is also clear that the complex time $s$ is translated through an imaginary displacement from the real time $t$. 

In the application of complex coordinates to express physical problems, there is generally going to be both a complex and a complex conjugate coordinate representation for each individual problem. In the present case, the complex conjugate of eq. (\ref{eq: conftrans1}) is
\begin{equation} \label{eq: conftrans2}
z_{i}^* = x_{i}, \quad s^* = t + \imath \frac{\hbar}{E} \left[ a\ln(r) + \left(\frac{r}{b} \right)^\lambda \right]
\end{equation}
Naturally, there must also be inverse transformations mapping the complex and complex conjugate representations of the problem back into a single physical coordinate system. The inverses of the transformations (\ref{eq: conftrans1}) and (\ref{eq: conftrans2}) are readily shown to be
\begin{equation} \label{eq: inv_ict1}
x_{i} = z_{i}, \quad t = s + \imath \frac{\hbar}{E} \left[ a\ln(r_z) + \left(\frac{r_z}{b} \right)^\lambda \right]
\end{equation}
\begin{equation} \label{eq: inv_ict2}
x_{i} = z_{i}^*, \quad t = s^* - \imath \frac{\hbar}{E} \left[ a\ln(r_z) + \left(\frac{r_z}{b} \right)^\lambda \right]
\end{equation}
($r_z =|z_i|$) respectively.  

It is now interesting to investigate properties of derivatives with respect to complex 4-position coordinates. In particular, the chain rule of partial differentiation gives
\begin{equation}  \label{eq: complexDiff1}
\frac{\partial}{\partial s}  
= \frac{\partial t}{\partial s} \frac{\partial}{\partial t}
+ \frac{\partial x_{i}}{\partial s} \frac{\partial}{\partial x_{i}}
= \frac{\partial}{\partial t} 
\end{equation}
\begin{equation} \label{eq: complexDiff2}
\frac{\partial}{\partial s^*}  
= \frac{\partial t}{\partial s^*} \frac{\partial}{\partial t}
+ \frac{\partial x_{i}}{\partial s^*} \frac{\partial}{\partial x_{i}}
= \frac{\partial}{\partial t} 
\end{equation}
\begin{equation} \label{eq: complexDiff3}
\frac{\partial}{\partial z_i} 
= \frac{\partial x_i}{\partial z_i} \frac{\partial}{\partial x_i}
+ \frac{\partial t}{\partial z_i} \frac{\partial}{\partial t}
= \frac{\partial}{\partial x_i} + \imath \frac{x_i}{r^2} \left[a + \lambda \left( \frac{r}{b} \right)^{\lambda} \right] \frac{\hbar}{E} \frac{\partial}{\partial t}
\end{equation}
\begin{equation} \label{eq: complexDiff4}
\frac{\partial}{\partial z_i^*} 
= \frac{\partial x_i}{\partial z_i^*} \frac{\partial}{\partial x_i}
+ \frac{\partial t}{\partial z_i^*} \frac{\partial}{\partial t}
= \frac{\partial}{\partial x_i} - \imath \frac{x_i}{r^2} \left[a + \lambda \left( \frac{r}{b} \right)^{\lambda} \right] \frac{\hbar}{E} \frac{\partial}{\partial t}
\end{equation}
Note, eqs. (\ref{eq: complexDiff1}) and (\ref{eq: complexDiff3}) have been obtained using eq. (\ref{eq: inv_ict1});  eqs. (\ref{eq: complexDiff2}) and (\ref{eq: complexDiff4}) are based on eq. (\ref{eq: inv_ict2}). It has also been assumed in deriving eqs. (\ref{eq: complexDiff1}) through (\ref{eq: complexDiff4}) that
\begin{equation} \label{eq: complexDiff5}
\frac{\partial z_i}{\partial s}=\frac{\partial s}{\partial z_i} = \frac{\partial z_i^*}{\partial s^*}=\frac{\partial s^*}{\partial z_i^*}=0
\end{equation}
indicating that the coordinates $z_i$ and $s$ are independent of each other as are the complex conjugate coordinates $z_i^*$ and $s^*$. This assumption is readily validated using eqs. (\ref{eq: complexDiff1}) through (\ref{eq: complexDiff4}) to directly evaluate each of the derivatives in eq. (\ref{eq: complexDiff5}) in $(x_i,t)$-coordinates. 

In further consideration of eqs. (\ref{eq: conftrans1}), it is convenient to write $s=t+i\tau$ where 
\begin{equation}
\tau = -\frac{\hbar}{E} \left[ a\ln(r) + \left(\frac{r}{b} \right)^\lambda \right]
\end{equation}
The requirement for a continuously differentiable function $f(s)=g(t,\tau)+ih(t,\tau)$ to be holomorphic is then for the real functions g and h to satisfy the set of Cauchy-Riemann equations
\begin{equation}
\frac{\partial g}{\partial \tau}=\frac{\partial h}{\partial t} \quad
\frac{\partial h}{\partial \tau}=-\frac{\partial g}{\partial t} 
\end{equation}
or equivalently
\begin{equation} \label{eq: creq}
\frac{\partial^2 f}{\partial t^2} + \frac{\partial^2 f}{\partial \tau^2} = 0
\end{equation}
It is thus concluded that a function $\psi(x_i,t)$ will also have an equivalent holomorphic form $\theta(z_i)f(s)$ in the complex $(z_\mu, s)$-coordinate system providing it is separable and $f$ satisfies eq. (\ref{eq: creq}). Here, it is understood that the domain of the Cauchy-Riemann equations in this problem is the complex plane containing s. The Cauchy-Riemann equations put no restriction at all on the form of the function $\theta(z_i)$ since $z_i$ and $s$ are independent coordinates and $z_i$ belongs to a real three-dimensional space.

\section{The Harmonic Oscillator}
The Klein-Gordon equation determining the wavefunction $\psi(x_i, t)$ for a single particle confined in a 3-dimensional harmonic oscillator potential can be expressed in the form
\begin{equation} \label{eq: KGHO1}
-\hbar^2c^2 \frac{\partial^2 \psi}{\partial x_i^2} + m_0^2c^4 \psi + \Omega^2 x^2\psi = E^2 \psi
\end{equation}
where $\Omega$ is the spring constant of the oscillator and
\begin{equation} \label{eq: KGHO2}
E\psi = \imath \hbar \frac{\partial \psi}{\partial t} 
\end{equation}
gives the total energy of the particle. 

The solution \cite{DFL} to eqs. (\ref{eq: KGHO1}) and (\ref{eq: KGHO2}) takes the separable form
\begin{eqnarray} \label{eq: hopsi1} 
\psi_{l_1 l_2 l_3}(x_i,t) = \phi_{l_1}(x_1)\phi_{l_2}(x_2)\phi_{l_3}(x_3)\exp(-\imath Et / \hbar)
\end{eqnarray}
where
\begin{eqnarray} \label{eq: phi1} 
\phi_l(x_i) =   k_l H_{l}(\xi_i) \exp \left(-\frac{\xi^2}{2} \right)
\end{eqnarray}
$\xi_i=\sqrt{\frac{\Omega}{\hbar}}x_i$, $H_{l_j}$ are Hermite polynomials and $l_1,l_2,l_3$ are positive integers. Inserting eq. (\ref{eq: hopsi1}) into eq. (\ref{eq: KGHO1}) gives the energy spectrum
\begin{equation}
E_n = \sqrt{2\hbar c \Omega \left(\frac{3}{2} + n \right)+ m_0^2c^4 }
\end{equation}
where $n=l_1 + l_2 + l_3$.

In developing the connection between complex $(z_i,s)$-coordinates and the quantum harmonic oscillator, the first step is to set $\lambda=2$ and
\begin{equation}
b=\sqrt{\frac{2\hbar c}{\Omega}}
\end{equation}
in eq. (\ref{eq: conftrans1}). In this case, eqs. (\ref{eq: complexDiff3}), (\ref{eq: complexDiff4}) and (\ref{eq: KGHO2}) can be combined to give
\begin{equation} \label{eq: complexDiff6}
\frac{\partial}{\partial z_i} 
= \frac{\partial}{\partial x_i} + \frac{\Omega}{\hbar c} x_i 
\end{equation}
\begin{equation} \label{eq: complexDiff7}
\frac{\partial}{\partial z_i^*} 
= \frac{\partial}{\partial x_i} - \frac{\Omega}{\hbar c} x_i
\end{equation}
These results lead to the operator relationship
\begin{equation}\label{eq: qprop1}
-\frac{\partial^2}{\partial z_i^* \partial z_i} + \frac{3 \Omega}{\hbar c}  = -\frac{\partial^2}{\partial x_i^2} + \frac{\Omega^2 x^2}{\hbar^2 c^2} 
\end{equation}
enabling the Klein-Gordon equation (\ref{eq: KGHO1}) for the harmonic oscillator to be expressed in the concise form
\begin{equation}\label{eq: complexKGHO1}
-\hbar^2 c^2 \frac{\partial^2 \psi}{\partial z_i^* \partial z_i}  + m_0^2c^4 \psi = \left( E^2 - 3\hbar c \Omega \right) \psi
\end{equation}
It is also readily shown using eqs. (\ref{eq: complexDiff1}) and (\ref{eq: KGHO2}) that
\begin{equation} \label{eq: complexKGHO2}
E\psi = \imath \hbar \frac{\partial \psi}{\partial s} 
\end{equation}
Eqs. (\ref{eq: complexKGHO1}) and (\ref{eq: complexKGHO2}) together, therefore, constitute a complete description of the quantum harmonic oscillator in terms of $(z_i,s)$-coordinates. On comparing eq. (\ref{eq: KGHO1}) and (\ref{eq: complexKGHO1}), it is clear that the harmonic oscillator potential term in the original Klein-Gordon equation is absent in the complex representation.

The oscillator function (\ref{eq: hopsi1}) is readily transformed into  $(z_i,s)$-coordinates using eqs. (\ref{eq: conftrans1}) to give
\begin{eqnarray} \label{eq: hopsi3} 
\psi(z_i,s) = \theta_{l_1}(z_1)\theta_{l_2}(z_2)\theta_{l_3}(z_3)f(s)
\end{eqnarray}
where
\begin{eqnarray} \label{eq: hopsi4} 
\theta_l(z_i) =  k_l H_{l}(\zeta_i), \quad f(s) = \exp(-\imath Es / \hbar)
\end{eqnarray}
and $\zeta_i = \sqrt{\frac{\Omega}{\hbar}}z_i$. It is notable that eq. (\ref{eq: hopsi1}) and (\ref{eq: hopsi3}) are similar except that eq. (\ref{eq: hopsi3}) does not contain a gaussian term. It is also notable that $f(s)$ is a continuously differentiable solution of eq.(\ref{eq: creq}) thus demonstrating that the oscillator function $\psi(z_i, s)$ is holomorphic.

It is instructive that the partial derivatives in $z_i$-space can be scaled to give the lowering $\hat{a}_\mu$ and raising $\hat{a}_\mu^\dag$ operators  
\begin{equation}  \label{ladder_nr}
\hat{a}_i = \sqrt{\frac{\hbar c}{2\Omega}}\frac{\partial}{\partial z_i}, \quad \hat{a}_i^\dag = -\sqrt{\frac{\hbar c}{2\Omega}}\frac{\partial}{\partial z_i^*}
\end{equation} 
Inserting these results into eq. (\ref{eq: complexKGHO1}) gives the expression
\begin{equation}
\hat{a}_i^\dag \hat{a}_i \psi = n \psi
\end{equation} 
It is clear therefore that the conformal transformation of the Klein-Gordon eq. (\ref{eq: KGHO1}) has led to the ladder operator representation of the harmonic oscillator. For an oscillator in the ground state $\psi_{000}$, eq. (\ref{eq: complexKGHO1}) simplifies to
\begin{equation}\label{eq: ground_state_ho}
\frac{\partial^2 \psi_{000}}{\partial z_i^* \partial z_i} = 0
\end{equation}
giving the solution $\psi_{000} = \exp(-\imath E_0s / \hbar)$. This result may also be cast into the familiar form $\hat{a}_i \psi_{000} = 0$. 

In consideration of the foregoing arguments, it is of interest that eqs. (\ref{eq: conftrans1}) reduces to the form $z_i=x_i, s=t$ on setting $\Omega=0$. It also apparent that eq. (\ref{eq: complexKGHO1}) reduces to the free field form of the Klein-Gordon equation under these same conditions. The converse of this argument is that harmonic interactions may be introduced into the free-field Klein-Gordon equation through the replacement $t \rightarrow t - \imath \frac{ \Omega}{2 E}x^2$ exactly equivalent to the more usual approach of adding the oscillator potential into the Hamiltonian for the oscillator.

\section{The Coulomb Potential}
The Klein-Gordon equation determining the wavefunction $\psi(x_i, t)$ for an electron of charge $-e$ bound in a Coulomb field originating from a fixed point charge $e$ can be expressed in the form
\begin{equation} \label{eq: KGCP1}
-\hbar^2c^2 \nabla^2 \psi + m_0^2c^4\psi = \left( E + c \hbar \frac{\alpha}{r} \right)^2 \psi
\end{equation}
where
\begin{equation}
\alpha = \frac{e^2}{4 \pi \epsilon_0 c \hbar} 
\end{equation}
is the fine structure constant and $\epsilon_0$ is the permittivity of free space.  This system is an approximation to the hydrogen atom neglecting the spin of the electron and the finite mass of the proton. Eq. (\ref{eq: KGHO2}) gives the total energy of the electron.

The solution \cite{DFL, JN} to eqs. (\ref{eq: KGCP1}) and (\ref{eq: KGHO2}) in spherical polar coordinates $(r,\theta,\phi)$ takes the separable form
\begin{eqnarray} \label{eq: psi_ha} 
\psi_{nlk}(r,\theta,\phi,t) = R_{nl}(r)Y_{lk}(\theta, \phi)\exp(-\imath E_nt / \hbar)
\end{eqnarray}
where $n,l,k$ are the hydrogenic quantum numbers,
\begin{equation}
R_{nl}(r) = \frac{\mathcal{N}_{nl}}{r^{\eta_l}} \exp \left( -\frac{r}{r_{nl}} \right)p_n\left(\frac{r}{r_{nl}}\right)
\end{equation}
$\mathcal{N}_{nl}$, $r_{nl}$ and $\eta_l$ are constants and $p_n$ is a polynomial of degree n. The normalized angular component $Y_{lk}(\theta, \phi)$ is unaffected by the conformal transformation. 

Inserting the wavefunction (\ref{eq: psi_ha}) into eq. (\ref{eq: KGCP1}) and collecting together terms in the same power of r gives
\begin{equation} \label{energy_ha}
E_{nl} = m_0c^2 \left[ 1 + \frac{\alpha^2}{(n+1-\eta_l)^2}\right]^{-1/2}
\end{equation}
\begin{equation}
r_{nl} = \frac{\hbar c (n+1-\eta_l)}{\alpha E_{nl}}
\end{equation}
where
\begin{equation}
\eta_l = \frac{1}{2} \pm \sqrt{\left(l+\frac{1}{2} \right)^2-\alpha^2}
\end{equation}
The expression for $\eta_l$ contains a $\pm$ sign. The negative sign is usually chosen since in this case eq. (\ref{energy_ha}) corresponds to the Sommerfeld energy spectrum for the hydrogen atom. By comparison, the positive sign predicts a much higher binding energy sometimes called the hydrino state.  

Setting $\lambda=1$ in eqs. (\ref{eq: complexDiff3}) and (\ref{eq: complexDiff4}) and making use of eq. (\ref{eq: KGHO2}) gives
\begin{equation} \label{eq: dz1_ha}
\frac{\partial}{\partial z_i} 
= \frac{\partial}{\partial x_i} + \frac{x_i}{r} \left(\frac{a}{r} + \frac{1}{b} \right)
\end{equation}
\begin{equation} \label{eq: dz2_ha}
\frac{\partial}{\partial z_i^*} 
= \frac{\partial}{\partial x_i} - \frac{x_i}{r} \left(\frac{a}{r} + \frac{1}{b} \right)
\end{equation}
such that
\begin{equation}\label{eq: d2z_ha}
\frac{\partial^2}{\partial z_i^* \partial z_i}  = \frac{\partial^2}{\partial x_i^2} + \frac{a}{r^2}(1-a) + \frac{2}{br}(1-a)-\frac{1}{b^2}
\end{equation}
Comparing the $r^{-1}$ and $r^{-2}$ terms in eqs. (\ref{eq: KGCP1}) and (\ref{eq: d2z_ha}) leads to the following explicit forms for the $a$ and $b$ coefficients  
\begin{equation}\label{eq: ct_coeffs_ha}
a = \eta_0, \quad b =\frac{\hbar c (1-\eta_0)}{\alpha E_{nl}}
\end{equation}
showing that $b$ depends on both the $n$ and $l$ quantum numbers. Eqs. (\ref{eq: ct_coeffs_ha}) enable eq. (\ref{eq: d2z_ha}) to be rewritten in the form of the operator relationship
\begin{equation}
\frac{\partial^2}{\partial z_i^* \partial z_i} + \frac{\alpha^2 E_{nl}^2}{\hbar^2 c^2 (1-\eta_0)^2}  = \frac{\partial^2}{\partial x_i^2} + \frac{\alpha^2}{r^2} + \frac{2E_{nl}}{\hbar c} \frac{\alpha}{r}
\end{equation}
Hence, using this last result the Klein-Gordon equation (\ref{eq: KGCP1}) for the hydrogen atom becomes
\begin{equation}\label{eq: complexKGCP1}
-\hbar^2 c^2 \frac{\partial \psi}{\partial z_i^* \partial z_i} + m_0^2 c^4 \psi = E_{nl}^2 \left[ 1 + \frac{\alpha^2}{(1-\eta_0)^2} \right] \psi 
\end{equation}
in the $(z_i,s)$-coordinates system. It is clear from inspection of this result that the goal of eliminating the Coulomb potential from eq. (\ref{eq: KGCP1}) has been achieved. Akin to eq. (\ref{eq: complexKGHO1}) for the harmonic oscillation, eq. (\ref{eq: complexKGCP1}) is similar in form to the free-particle Klein-Gordon equation except that the eigenvalues are a different function of the total energy. 

The hydrogenic wave function (\ref{eq: psi_ha}) is readily transformed into the spherical polar form of $(z_i,s)$-coordinates using eqs. (\ref{eq: conftrans1}) to give
\begin{eqnarray} \label{eq: complex_psi_ha} 
\psi_{nlk}(r_z,\theta_z,\phi_z,s) = R_{nl}(r_z)Y_{lk}(\theta_z, \phi_z)\exp(-\imath Es / \hbar) 
\end{eqnarray}
where
\begin{equation}\label{eq: Rz}
R_{nl}(r_z) =  \mathcal{N}_{nl}\exp \left[ -\frac{(n-\eta_l+\eta_0)}{(1-\eta_0)} \frac{r_z}{r_{nl}}\right] \frac{r_z^{\eta_0}}{r_z^{\eta_l}} p_n\left( \frac{r_z}{r_{nl}} \right)
\end{equation}  
Here, $\exp(-\imath Es / \hbar)$ is a continuously differentiable solution of eq.(\ref{eq: creq}) thus demonstrating that eq. (\ref{eq: complex_psi_ha}) is holomorphic.

For the ground state of the hydrogen atom $\psi_{000}$, eq. (\ref{eq: complexKGCP1}) reduces to eq. (\ref{eq: ground_state_ho}) giving the solution $\psi_{000} = \exp(-\imath E_0s / \hbar)$ identical to the harmonic oscillator.

\section{Concluding Remarks}
It has been shown the concept of a potential field can be eliminated from the mathematical description of both the relativistic quantum harmonic oscillator and the hydrogen atom through the use of an isometric conformal transformation. In the transformed coordinate system, time is a complex quantity. The real part of this complex time is the world time; the imaginary part is responsible for binding the particles into their respective systems.

\end{document}